\documentstyle[twoside,fleqn,espcrc2,epsfig]{article}
\input {epsf.sty}
\input {psfig.sty}

\newcommand{\AmS}{{\protect\the\textfont2
  A\kern-.1667em\lower.5ex\hbox{M}\kern-.125emS}}
\newcommand{\gapprox}{\raisebox{-0.5ex}{$\
\stackrel{\textstyle>}{\textstyle\sim}\ $}}
\newcommand{\lapprox}{\raisebox{-0.5ex}{$\
\stackrel{\textstyle<}{\textstyle\sim}\ $}}

\title{Lattice Matter}

\author{Simon Hands\address{Department of Physics,
        University of Wales Swansea, \\ 
        Singleton Park, Swansea SA2 8PP, United Kingdom}%
        }
       
\begin{document}

\begin{abstract}
I review recent developments in the study of strongly interacting field theories
with non-zero chemical potential $\mu$. In particular I focus on ({\em a\/}) the
determination of the QCD critical endpoint in the $(\mu,T)$ plane; ({\em b\/})
superfluid condensates in Two Color QCD; and ({\em c\/}) Fermi surface effects 
in the NJL model. Some remarks are made concerning the relation of
superconductivity with the sign problem.
\end{abstract}

\maketitle

\begin{figure}[htb]
\centerline{
\setlength\epsfxsize{230pt}
\epsfbox{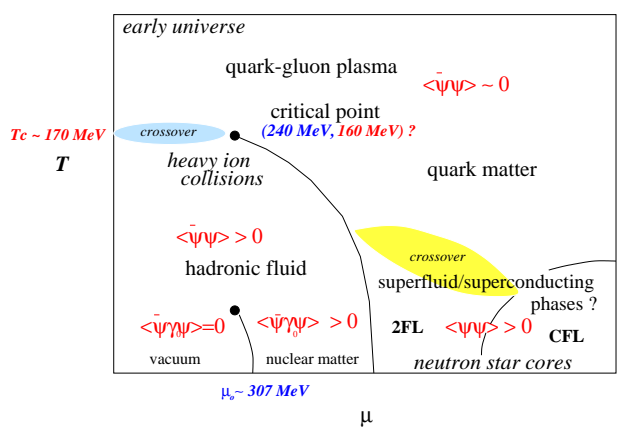}
}
\caption{Schematic view of the QCD phase diagram.}
\label{fig:qcd}
\end{figure}
Fig.~\ref{fig:qcd} summarises our current knowledge of the QCD phase diagram
in the plane of temperature $T$ and quark chemical potential $\mu$. Last year 
Shinji Ejiri \cite{Shinji} reviewed QCD simulations at $T\not=0$, ie. along the 
vertical axis. In my talk I wish to discuss what can be done in the interior of
the plane. 
Significant progress has been made in the region to the 
upper left of Fig.~\ref{fig:qcd} where both $\mu$ and $T$ differ from zero,
which is also of direct phenomenological interest for heavy-ion
collisions. I also wish to cover the lower right region describing cold dense
strongly-interacting matter.
In recent years there has been intense theoretical activity in this region
driven by the possibility that quark matter is unstable with 
respect to {\sl diquark condensation\/} $\langle qq\rangle\not=0$, resulting in 
a ground state with color superconducting properties
\cite{super}.
Model calculations suggest that the BCS gap $\Delta$ at the Fermi surface 
may be as large as 100 MeV \cite{BR}, 
comparable with the constituent quark scale,
implying significant consequences for the physics of compact astrophysical
objects \cite{ABR}.
In this case simulations of QCD are to date impracticable, for reasons I will
review below; however, diquark condensation and Fermi surface effects can be
studied by lattice techniques in certain models.

\section{Why is $\mu\not=0$ so difficult?}
\label{sec:mu}

For a vectorlike gauge theory in Euclidean metric 
the introduction of a quark chemical potential
breaks the $\gamma_5$-hermiticity of the Dirac operator:
\begin{equation}
D{\!\!\!\! /\,}(\mu)\equiv D{\!\!\!\! /\,}(0)+\mu\gamma_0=
\gamma_5D{\!\!\!\! /\,}^\dagger(-\mu)\gamma_5.
\end{equation}
This implies that eigenvalues of $D{\!\!\!\! /\,}$ are no longer pure
imaginary and hence not related to each other by complex conjugation,
in turn implying 
\begin{equation}
\mbox{det}M(\mu)\not=\mbox{det}M^*(\mu)=\mbox{det}M(-\mu).
\end{equation}
Therefore the functional measure is no longer positive definite.
In principle the determinant
can be factorised into a modulus $\rho$ and a phase $\phi$, 
and the phase included with the 
observable $\cal{O}$ in Monte Carlo simulations via
\begin{equation}
\langle{\cal O}\rangle\equiv\langle{\cal O}e^{i\phi}\rangle_\rho/
\langle e^{i\phi}\rangle_\rho.
\end{equation}
Unfortunately fairly general arguments suggest that 
$\langle e^{i\phi}\rangle_\rho\propto e^{-V}$, where $V$ is the system volume.
Acquiring sufficient statistics therefore becomes exponentially difficult as the
thermodynamic limit is approached. This is known as the {\sl Sign Problem\/},
and has plagued the study of $\mu\not=0$ since its inception. However, 
we should not be so surprised, since generic problems in the quantum theory 
of $N$ objects require $N!\sim e^N$ different complexions or wavefunctions
to be examined. Perhaps a more productive way of phrasing the question 
would be {\sl why is vacuum QCD so easy?\/}

Given this difficulty there are two routes forward. Firstly, one can perform a
QCD simulation at $\mu=0$ and attempt to analytically continue the results to
$\mu\not=0$. This can be done either by calculating terms in a Taylor expansion
about $\mu=0$, exemplified, for instance, by the calculation of 
baryon number susceptibility $\chi_B=V^{-1}\partial^2\ln
Z/\partial\mu^2\vert_{\mu=0}$
\cite{Gottlieb},
or by directly reweighting configurations, as in the `Glasgow
method' developed by Ian Barbour and collaborators \cite{Barbour}.
In the former case the prospects are limited by a finite radius of convergence,
dictated by the presence of a critical point in the $(\mu,T)$ plane. In the
latter case the effectiveness is limited by the requirement of maintaining 
a reasonable overlap between the trial and true ensembles, which once again
becomes exponentially hard in the thermodynamic limit \cite{Halasz}. 
Both methods have
their best chance of succeeding at $T>0$; at $T=0$ the behaviour is
horribly non-analytic, since the ground state does not change as $\mu$ 
increases out to the onset
of nuclear matter at $\mu_o\simeq307$MeV. As we shall see, the overlap problem 
is also considerably more tractable at $T>0$.

The second approach is to throw up our hands and use a real measure
$\mbox{det}MM^*=\mbox{det}M(\mu)M(-\mu)$. Physically this has the 
effect of introducing conjugate quarks $q^c$ which have positive baryon charge 
but transform in the conjugate representation of the gauge group \cite{qc},
leading
to the possibility of gauge invariant $qq^c$ bound states. In real QCD this is a
disaster since the lightest such state is degenerate with the pion, being in
effect a `Goldstone baryon'. As $\mu$ is raised the onset of nuclear matter 
would thus be expected at $\mu_0\approx m_\pi/2$ rather than the constituent
quark scale $\Sigma\approx m_N/3$. There are models, however, such as Two
Color QCD \cite{Elbio} and QCD with non-zero isospin chemical potential
$\mu_I\propto\mu_u-\mu_d$ \cite{SS},
where $qq$ Goldstone baryons are a feature, not a problem \cite{KSTVZ}. 
Another class 
amenable to this approach are NJL-like models, where $qq^c$ states don't couple
to the Goldstone mode \cite{BHKLM} and hence remain at the constituent scale, 
so the use of a real measure doesn't
damage the physics. The common feature of all these models is that the diquark
condensate which potentially forms at large $\mu$ is gauge singlet, and hence
describes a superfluid, as opposed to superconducting, ground state.

\section{Progress at $T>0$}

Fodor and Katz \cite{FK} have made spectacular progress in reweighting
at $T>0$. The basic formula is
\begin{eqnarray*}
Z[\alpha]=\int\!DU
\exp(-S_{bos}[U;\alpha_0])\mbox{det}M[U;\alpha_0]\times\nonumber\\
\left\{\exp(-\Delta S_{bos}[U;\alpha,\alpha_0]){{\mbox{det}M[U;\alpha]}\over
{\mbox{det}M[U;\alpha_0]}}\right\}
\end{eqnarray*}
where the parameter set $\alpha=\{\beta,m,\mu\}$ but importance sampling 
is performed using 
a different set $\alpha_0=\alpha-\Delta\alpha$,
with $\mu$ chosen either zero or pure imaginary to keep the determinant real. 
Usually
reweighting is only effective if $\Delta\alpha$ is small enough to maintain a
good overlap between trial and true ensembles. Fodor and Katz' insight is that 
along the crossover/coexistence line between 
hadronic and quark-gluon plasma phases the true ensemble 
cannot alter very much,
since it must contain
contributions from both phases; hence the overlap between ensembles along
the line remains high. 
They
therefore generalise the Glasgow method by 
reweighting with both $\Delta\mu,\Delta\beta\not=0$ in order to stay on
this line. In practice $\beta_c(\mu)$ is identified via the real part of the 
lowest Lee-Yang zero $\beta_0$, which approaches the real axis as $V$ increases.
Now, the nature of the transition between hadronic and QGP phases can be
determined by the volume scaling of $\mbox{Im}\beta_0$:
\begin{equation}
\lim_{V\to\infty}\mbox{Im}\beta_0\cases{\not=0;&crossover,\cr
           =0;&first order transition.}
\end{equation}

\begin{figure}[htb]
\centerline{
\setlength\epsfxsize{230pt}
\epsfbox{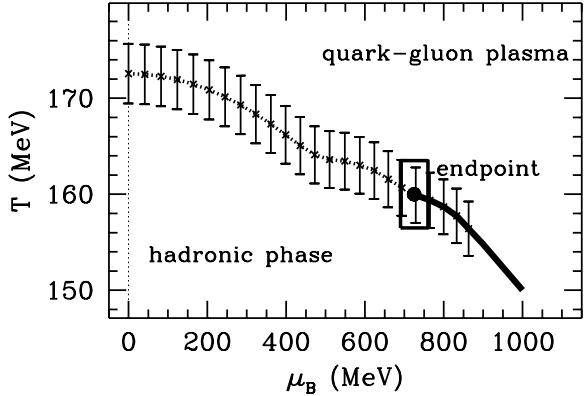}
}
\caption{Reweighting prediction for 
the quark-hadron transition (note $\mu_B\equiv3\mu$).}
\label{fig:LY}
\end{figure}
The results for $2+1$ flavor QCD with $m_{u,d}a=0.025$, $m_sa=0.2$ on volumes up
to $8^3\times4$ are shown in Fig.~\ref{fig:LY}. The physical scale is set using
separate simulations at $\beta_c(0)$ and $\beta_c(\mu_E)$, which yield a
critical temperature $T_c(\mu=0)=172(3)$MeV. 
The main prediction is for the 
critical endpoint of the first order line, at
\begin{equation}
T_E=160(4)\,\mbox{MeV}\;\;;\;\;\mu_E=242(12)\,\mbox{MeV}.
\label{eq:endpoint}
\end{equation}
The result for $\mu_E$ is unexpectedly large, and might undermine proposals to
observe it directly at RHIC \cite{SRS}. However, the endpoint is expected to
move to the left as the chiral limit is approached, so one should consider
(\ref{eq:endpoint}) more as a proof of principle at this stage. What should be
stressed is that currently the results are limited by cpu resources 
rather than overlap problems as $V\to\infty$; it is certainly possible,
therefore, that lattice methods can ultimately obtain an accurate determination
of $(\mu_E,T_E)$.

There have also been developments in Taylor expansion methods, which can deliver
information on a
wider range of observables. Gavai and Gupta have reexamined quark number
susceptibilities in quenched QCD \cite{GG}, in particular finding 
that $\chi_s$ increases markedly across the transition to QGP, and approaches
$\chi_{u,d}$ for $T\gapprox2T_c$, though all three depart significantly from
ideal-gas values. These quantities are testable in ion collisions via eg.
event-by-event fluctuations in charged particle yields and strangeness
enhancement. QCD-TARO have looked
at the response of hadron masses to a change of both baryon and isospin
chemical potentials \cite{QCDTARO}. They
find $\partial^2m_\pi/\partial\mu_B^2>0$, and with a much larger value in the
QGP phase, indicative that the pion is no longer a Goldstone mode at high $T$;
by contrast $\partial^2m_\pi/\partial\mu_I^2<0$ and is much larger in the
hadronic phase, consistent with pion condensation in cold dense
isospin-asymmetric matter \cite{SS}. Finally Ejiri has determined the
curvature of the critical line $\partial^2\beta_c(\mu)/\partial\mu^2$ by 
measuring the shift in the peaks for $\langle\bar\psi\psi\rangle$ and Polyakov
loop susceptibities \cite{Shinji2}. For quark mass $ma=0.2$ on $16^3\times4$
he obtains a value $\simeq-1.5(4)$; assuming a critical temperature 
$T_c\approx170$MeV and a 
$\beta$-function given by its one-loop value, this
yields a value $T_c(\mu\!=\!242\mbox{MeV})\!\approx150\,\mbox{MeV}$, 
not too far 
from (\ref{eq:endpoint}).

\section{Two colors matter}

For gauge group SU(2), the fermion determinant is trivially real since all
matter representations are either real, or pseudoreal (in which case
$\mbox{det}M=\mbox{det}\tau_2M^*\tau_2$). Moreover, if we specialise to the case
of $N$ staggered fermions in the fundamental representation, the usual 
$\mbox{U}(N)\otimes\mbox{U}(N)_\varepsilon$ global
symmetry at $m=\mu=0$ is enhanced,  
via 
\begin{eqnarray}
\bar\chi D{\!\!\!\!/\,}\chi&=&\bar X_eD{\!\!\!\!/\,}X_o,\nonumber\\
\bar X_e=(\bar\chi_e,-\chi_e^{tr}\tau_2) &;&
X_o=\left(\matrix{\chi_o\cr-\tau_2\bar\chi_o^{tr}}\right)
\label{eq:U2N}
\end{eqnarray}
to $X\mapsto VX$, $\bar X\mapsto\bar XV^\dagger$, $V\in\mbox{U}(2N)$.
For adjoint quarks the same relation holds with the $\tau_2$ factors 
replaced by 1. Now, at $\mu=0$ we expect this global symmetry to be
spontaneously broken by a chiral condensate $\langle\bar\chi\chi\rangle\not=0$.
For fundamental quarks the breaking pattern is $\mbox{U}(2N)\to\mbox{O}(2N)$
yielding $N(2N+1)$ Goldstones; for adjoint the pattern is 
$\mbox{U}(2N)\to\mbox{Sp}(2N)$ with $N(2N-1)$ Goldstones \cite{HMMOSS}. This 
differs in detail from the corresponding patterns for continuum fermions
\cite{KSTVZ}. The important point is that besides the usual
mesonic $q\bar q$
states, in general some of the Goldstones 
must be $qq$ baryons, as anticipated in Sec.~\ref{sec:mu}. It is also possible
to identify diquark condensates for both {\bf2} and {\bf3} representations:
\begin{equation}
\left\langle\left\{\matrix{qq_{\bf2}\cr qq_{\bf3}}\right\}\right\rangle
={1\over2}\left\langle
\chi^{tr}\left\{\matrix{\tau_2\cr i\epsilon}\right\}\chi+
\bar\chi\left\{\matrix{\tau_2\cr i\epsilon}\right\}\bar\chi^{tr}
\right\rangle
\label{eq:qq}
\end{equation}
where in the {\bf3} case the antisymmetric $\epsilon$ acts on flavor. These
are related to $\langle\bar\chi\chi\rangle$ by a $\mbox{U}(2N)$ rotation and
are gauge invariant.
Since, however, (\ref{eq:qq}) is not invariant under the original U(1)$_B$,
baryon charge is no longer a good quantum number.

It is possible to treat TCQCD analytically by ignoring all excitations except
the Goldstones, and writing down an effective action in the spirit of chiral
perturbation theory ($\chi$PT), in which the physical parameters
are $\langle\bar\chi\chi\rangle$ and $m_\pi$ at $\mu=0$. Remarkably, the 
$\mu$-dependent terms in the chiral Lagrangian are completely determined
in terms of these two parameters by the global symmetries \cite{KSTVZ}.
At leading order in $\chi$PT a second order onset phase transition is predicted
at the rescaled chemical potential 
$x\equiv2\mu/m_\pi=1$:
\begin{eqnarray}
{{\langle\bar\chi\chi\rangle}\over{\langle\bar\chi\chi\rangle_0}}=
\cases{1\cr{1\over x^2}};\;\;\;
{{\langle qq\rangle}\over{\langle\bar\chi\chi\rangle_0}}
=\cases{0\cr \sqrt{1-{1\over{x^4}}}};
\nonumber\\
{N\over{2m\langle\bar\chi\chi\rangle_0}}\,n_B=
\cases{0;&$x<1$\cr 
x\left(1-{1\over x^4}\right);&$x>1$}
\label{eq:chiPT}
\end{eqnarray}
where the baryon charge density
$n_B=\langle\bar\psi\gamma_0\psi\rangle$ 
and the 0 subscript denotes
values at $\mu=0$. The high-$\mu$ phase is a superfluid which forms as a result
of a 
Bose-Einstein condensation of weakly interacting
diquark bosons.

Let us briefly discuss the measurement of $\langle qq\rangle$. In a 
finite volume
the technicalities are
identical to those involved in measuring the chiral condensate.
A diquark source
term $j\chi^{tr}\tau_2\chi$ is introduced and the action rewritten in a Gor'kov
basis \cite{MH}:
\begin{equation}
{\cal L}=(\bar\chi,\chi^{tr})\left
(\matrix{\!\!\bar\jmath\tau_2&\!\!{1\over2}M\cr
\!\!-{1\over2}M^{tr}&\!\!j\tau_2}\right)
\left(\matrix{\!\!\bar\chi^{tr}\cr\!\!\chi}\right)
\equiv\Psi^{tr}{\cal A}\Psi
\nonumber
\end{equation}
whence
\begin{equation}
Z[j,\bar\jmath]=\int{\cal D}U\mbox{Pf}(2{\cal A}[U,j,\bar\jmath])
e^{-S_{bos}[U]}.
\label{eq:pf}
\end{equation}
The condensate is then defined by
\begin{equation}
\langle qq(j)\rangle={1\over V}{{\partial\ln Z}\over{\partial j}}=
{1\over{2V}}\langle\mbox{tr}\tau_2{\cal A}^{-1}\rangle.
\end{equation}
Of course, since $j$ is not a physical parameter 
one wants the $j\to0$ limit, but any favourite method for probing
$\langle\bar\psi\psi(m=0)\rangle$ can be used, such as direct inversion
of ${\cal A}(j)$ followed by extrapolating $j\to0$ \cite{KSHM}, using a
Banks-Casher relation on the eigenvalue density of $\tau_2{\cal A}$ \cite{BLMP},
or by calculating the probability distribution function of $\langle qq\rangle$
\cite{AADGG}. 

\begin{figure}[htb]
\centerline{
\epsfig{file=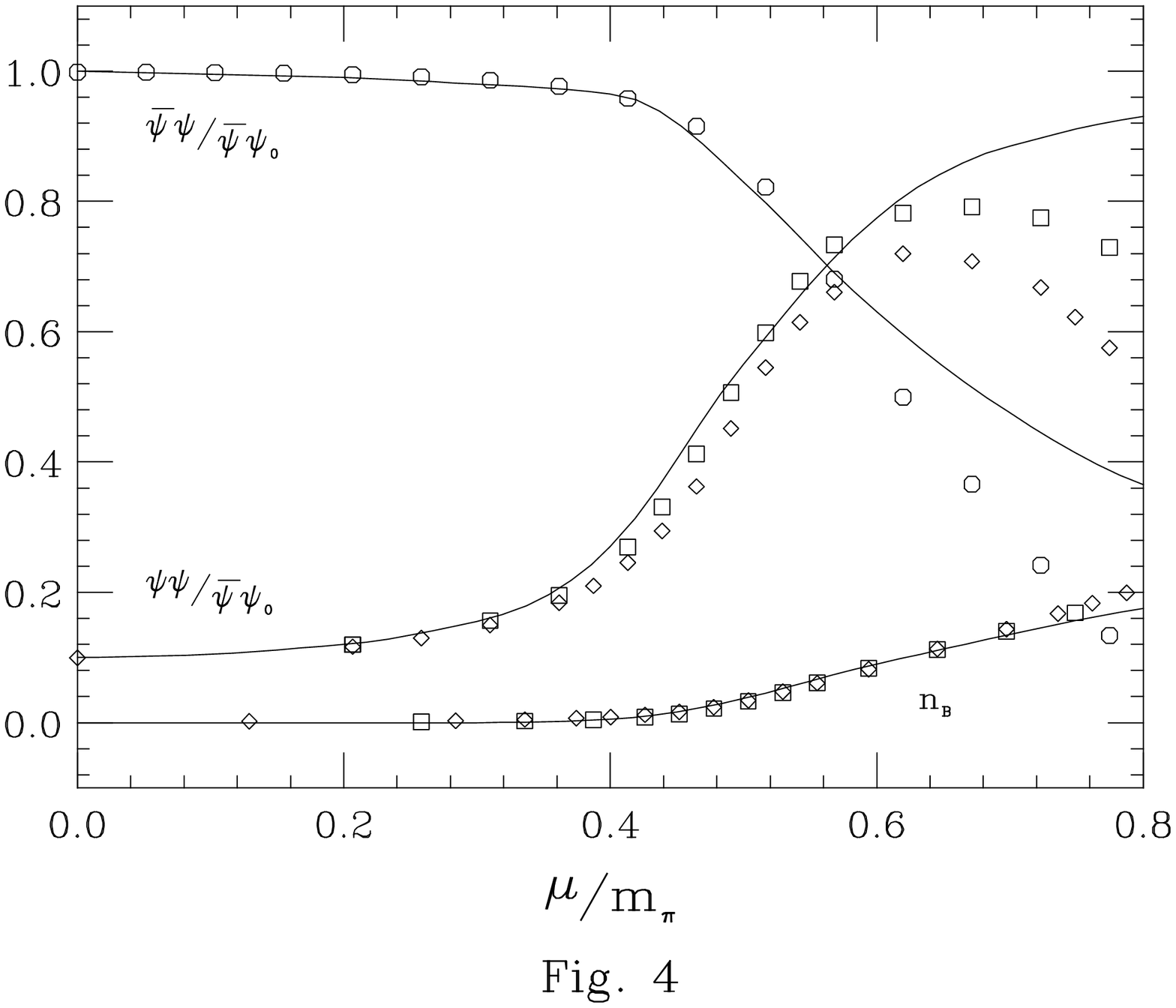,width=200pt,angle=90,clip=}
}
\caption{$\langle\bar\chi\chi\rangle$, $\langle qq\rangle$ and $n_B$ vs. $\mu$
for $\beta=0$, $ma=0.2$, $j/m=0.1$ on $4^4$ and $6^4$.}
\label{fig:azcoiti}
\end{figure}

\begin{figure}[htb]
\centerline{
\setlength\epsfxsize{230pt}
\epsfbox{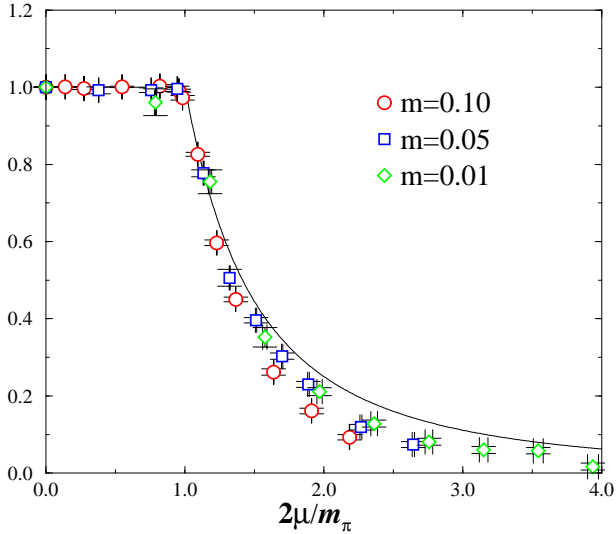}
}
\caption{$\langle\bar\chi\chi\rangle/\langle\bar\chi\chi\rangle_0$ vs. $x$
for $\beta=2.0$, $ma=0.1,0.05$ and 0.01 on $4^3\times8$.}
\label{fig:unipbp}
\end{figure}

\begin{figure}[htb]
\centerline{
\setlength\epsfxsize{230pt}
\epsfbox{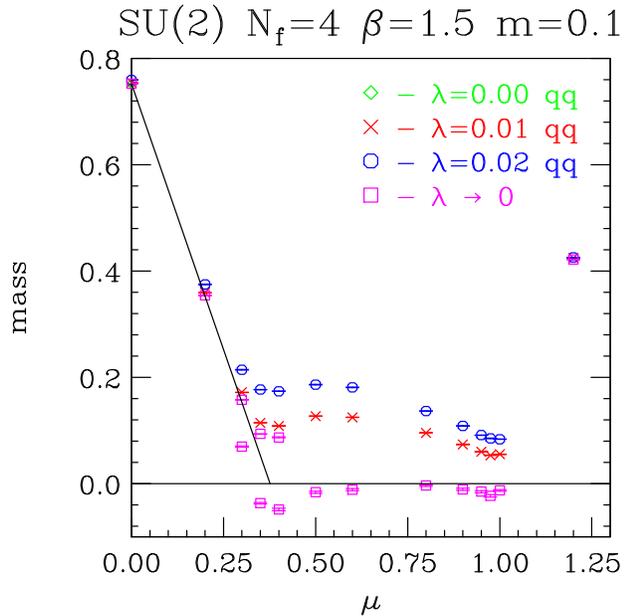}
}
\vskip -25pt
\caption{Scalar diquark mass vs. $\mu$ for various diquark source strengths
$\lambda$ on $12^3\times24$.}
\label{fig:Goldstone}
\end{figure}

Since the pioneering work of Nakamura \cite{Nak}, there have been several
studies of TCQCD with $\mu\not=0$ 
using a variety of algorithms and actions
\cite{Elbio,HMMOSS,MH,KSHM,BLMP,AADGG,rest,HMSS,KTS}. I will briefly discuss
some highlights. 
Fig.~\ref{fig:azcoiti} shows strong coupling results \cite{AADGG} for 
$\langle\bar\chi\chi\rangle$, $n_B$ and $\langle
qq\rangle$ vs. $\mu$ together with $\chi$PT predictions (\ref{eq:chiPT}), 
showing acceptable agreement until saturation artifacts creep in for 
$\mu/m_\pi\gapprox0.6$.
Fig.~\ref{fig:unipbp} shows
$\langle\bar\chi\chi\rangle/\langle\bar\chi\chi\rangle_0$ vs. rescaled
chemical potential for three different bare quark masses in TCQCD with
adjoint quarks \cite{HMMOSS,HMSS}, together 
with the prediction (\ref{eq:chiPT}).
In this case
$\chi$PT appears to work well up to $x\approx2$ over a decade of quark mass.
Since $\langle qq\rangle\not=0$ breaks a global symmetry, we anticipate a 
diquark Goldstone boson for $\mu>\mu_o$ 
whose mass should vanish as $j\to0$ \cite{KSTVZ}. Physically this results in
long-ranged interactions between vortex excitations in a rotating superfluid,
and in propagating waves of temperature variation known as {\sl second sound\/}.
Simulations with $j=0$ slow dramatically in this regime due to the 
profusion of small eigenvalues of $M$ \cite{HMMOSS}. Fig.~\ref{fig:Goldstone}
shows the results of simulating the Pfaffian weight (\ref{eq:pf})
with $j\not=0$ on a
much larger system \cite{KSHM}, which enable the identification of a massless
mode in the zero source limit.
Finally, studies of the high density regime at $T>0$ have shown a strong
first-order transition restoring the normal QGP state $\langle qq\rangle=0$
\cite{KTS}.

\begin{figure}[htb]
\centerline{
\epsfig{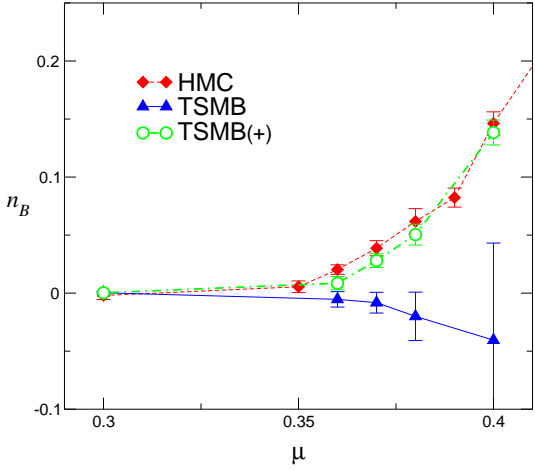}
}
\caption{Baryon charge density $n_B$ vs. $\mu$ in adjoint TCQCD 
showing that sign reweighting removes the transition observed in the
$\mbox{det}>0$
sector.
}
\label{fig:tsmb}
\end{figure}

Although the TCQCD measure is obviously real, we have not discussed its
positivity. In fact, it is possible to prove it positive for all cases
except an odd number of adjoint staggered quarks. For $N=1$
adjoint flavor, however, the sign problem returns \cite{HMMOSS}. Additionally
in this case,
the Goldstone count does not include any baryons, and the superfluid condensate
$\langle qq_{\bf3}\rangle$ in (\ref{eq:qq}) is forbidden by the Pauli Exclusion
Principle. Both facts invalidate the use of $\chi$PT. Indeed, the simplest 
local diquark operator that can be written is gauge-variant; 
\begin{equation}
qq^i_{sc}={1\over2}\left[\chi^{tr}t^i\chi+\bar\chi t^i\bar\chi^{tr}\right]\in
{\bf3}\;\; \mbox{of  SU}(2),
\end{equation}
leading to the possibility of color superconductivity {\em\`a la\/}
Georgi-Glashow in the high density phase.
Simulations using the two-step multibosonic algorithm have been performed which
reweight the ensemble taking the sign into account \cite{HMMOSS}.
Once $\mu>\mu_o$ the sign starts to fluctuate, and by $\mu=0.38$ at $m=0.1$
$\langle\mbox{sgn}(\mbox{det}M)\rangle$ has fallen to 0.30(4) on $4^3\times8$ 
\cite{HMSS}. Remarkably, once the observables are also reweighted, all signals
for the onset transition disappear, as seen in Fig.~\ref{fig:tsmb}.
This is in accord with the expectation that there is now a separation 
between the lightest baryon mass and the Goldstone scale, and demonstrates that
sgn(det) plays a decisive role in determining the ground state.
Fig.~\ref{fig:tsmb} is probably the most expensive simulation of nothing
happening to date.

\section{Flatland NJL}

The Nambu -- Jona-Lasinio model has a long history as an effective description
of the strong interaction, and also has thermodynamic applications
\cite{Klevansky}. The Lagrangian is
\begin{equation}
{\cal L}=\bar\psi(\partial{\!\!\!/\,}+\mu\gamma_0+m)\psi-{g^2\over2}\left[
(\bar\psi\psi)^2-(\bar\psi\gamma_5\vec\tau\psi)^2\right]
\end{equation}
At zero chemical potential and for 
$m=0$ it has a \hbox{SU(2)$_L\otimes$SU(2)$_R\otimes$U(1)$_B$}
global symmetry, which for sufficiently strong coupling 
$g^2\geq g_c^2$ spontaneously  breaks
to SU(2)$_I\otimes$U(1)$_B$ accompanied by the dynamical generation of a
constituent mass $\Sigma=g^2\langle\bar\psi\psi\rangle$. Our current interest is
the model in 2+1 dimensions, since apart from the obvious computational saving
there is an interacting continuum limit at $g^2\to g_c^2$, $\Sigma a\to0$
\cite{RWP}. For $\mu\not=0$, there is a strong first-order chiral
symmetry-restoring transition at $\mu_c\simeq\Sigma$ \cite{Njlmu}, which is 
completely separate from the Goldstone scale \cite{BHKLM,HKK}, but appears
to
coincide \cite{Strouthos} with an onset transition 
separating the vacuum from a regime with 
$n_B\propto\mu^2$, suggestive of a two-dimensional Fermi surface with 
$E_F\simeq\mu$ in the chirally-restored phase. This is natural since the 
lightest baryons in the model are fermions.

\begin{figure}[htb]
\centerline{
\setlength\epsfxsize{230pt}
\epsfbox{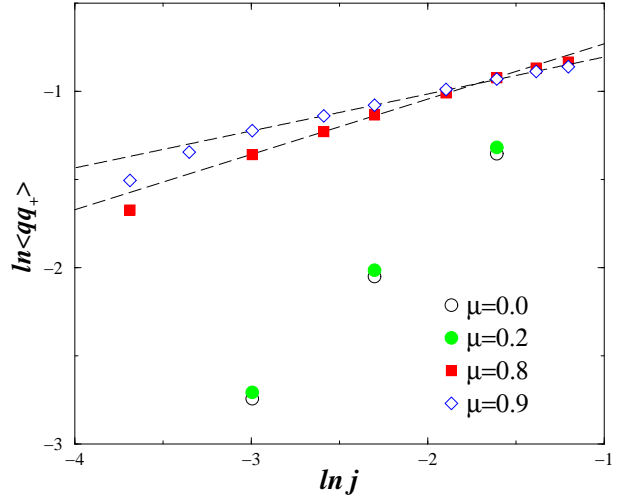}
}
\vskip -25pt
\caption{$\ln\langle qq_+\rangle$ vs. $\ln j$ for $\mu=0.0,0.2,0.8,0.9$. }
\label{fig:lnln}
\end{figure}

\begin{figure}[htb]
\centerline{
\setlength\epsfxsize{230pt}
\epsfbox{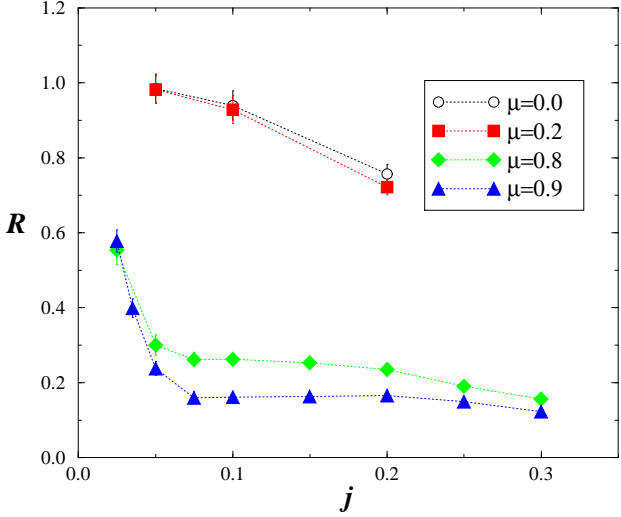}
}
\vskip -25pt
\caption{$R(j)$ extrapolated to $L_t=\infty$ vs. $j$.
}
\label{fig:R}
\end{figure}

An obvious question is whether U(1)$_B$ is spontaneously broken for $\mu>\mu_c$
by a superfluid condensate $\langle\chi^{tr}\tau_2\chi\rangle\not=0$
(note that we
have surreptitiously slipped back into the notation of staggered fermions;
the continuum translation is given in \cite{HM}).
Superfluid condensation in this model would occur via a BCS instability at the
Fermi surface, as in $^3$He. To investigate this we have performed Pfaffian 
simulations with diquark source term \cite{HLM}
\begin{equation}
j_\pm qq_\pm=j_\pm(\chi^{tr}\tau_2\chi\pm\bar\chi\tau_2\bar\chi^{tr}),
\end{equation}
measuring both condensate $\langle qq_+\rangle$ and associated
susceptibilities $\chi_\pm=\sum_x\langle qq_\pm(0)qq_\pm(x)\rangle$. 
Analogous to the axial Ward identity, we have
\begin{equation}
\label{eq:ward}
\chi_-(j_-=0)={{\langle qq_+\rangle}\over{j_+}}.
\end{equation}
The results are unexpected: fig.~\ref{fig:lnln} shows a log-log plot of 
$\langle qq_+\rangle$ vs. $j_+$ from data from volumes up to $32^3$ extrapolated
to the zero temperature limit $L_t\to\infty$. The high density
data ($\mu_c\simeq0.65$ for our choice of coupling) suggest a power-law
equation of state
\begin{equation}
\langle qq_+\rangle\propto j^\alpha,
\label{eq:powerlaw}
\end{equation}
with $\alpha=\alpha(\mu)$ 
falling in the range 0.2 - 0.3 for the values of $\mu$ examined. This is
reinforced by the susceptibility ratio $R=\vert\chi_+/\chi_-\vert$ plotted in 
fig.~\ref{fig:R}; using (\ref{eq:ward}) and (\ref{eq:powerlaw}) it is easy
to show that $R(j)$ should take the constant value $\alpha$. The plateaux
in fig.~\ref{fig:R} display this behaviour, with approximately the same 
$\alpha(\mu)$
as those from  direct fits to (\ref{eq:powerlaw}).

The strange behaviour of figs.~\ref{fig:lnln},\ref{fig:R} is strongly
reminiscent of the low temperature $2d$ $XY$ model, which exists in a critical
state for a range of $T$ with continuously varying exponents $\delta(T)$,
$\eta(T)$, which we can define analogously (the arrow denotes a $2d$ vector):
\begin{equation}
\langle qq\rangle\propto j^{1\over\delta}\;\;;\;\;
\langle qq(0)qq(\vec x)\rangle\propto{1\over{\vert\vec x\vert^\eta}}.
\label{eq:XY}
\end{equation}
We therefore conjecture that NJL$_{2+1}$ in its large-$\mu$ phase 
describes
a $2d$ critical system. The condensate is washed out by long-wavelength
fluctuations as $j\to0$, but long range phase coherence is maintained via
(\ref{eq:XY}). Intriguingly, the most precise tests 
of the $2d$ $XY$ universality class are from experiments performed on thin 
films of superfluid $^4$He \cite{Nelson}. The superfluid current is related 
to the phase $\theta(x)$ of $qq(x)$  via
\begin{equation}
\vec J_s=K_s\vec\nabla\theta.
\end{equation}
There is a beautiful topological argument \cite{KT} that the only way to change
the circulation $\kappa=\oint \vec J_s.\vec dl$ around a periodic volume is to 
create a vortex -- anti-vortex pair and translate one of them around the
universe in the perpendicular direction until they reannihilate, 
thereby increasing 
$\kappa$  by a quantum
$2\pi K_s/L$. Since, however, the energy required increases logarithmically with
pair separation, the circulation is metastable, thereby
demonstrating superfluidity.

\begin{figure}[htb]
\centerline{
\setlength\epsfxsize{230pt}
\epsfbox{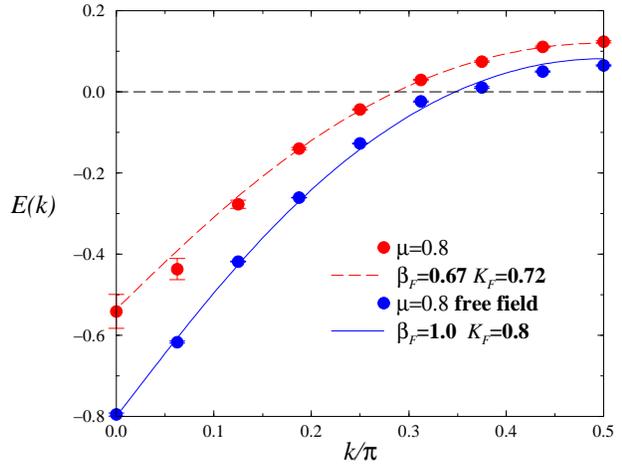}
}
\vskip -25pt
\caption{$E(k)$ for both free and interacting quasiparticles
at $\mu=0.8$.
}
\label{fig:quasi}
\end{figure}

Why, then, does the NJL exponent $\delta(\mu)\approx3 - 5$ differ from the 
$XY$ value $\delta(T)\geq15$? Standard dimensional reduction 
does not apply since the limit $L_t\to\infty$ is needed, implying that the 
fermion modes do not decouple. Further insight is gained from studying the 
fermion spectrum via the Gor'kov propagator ${\cal G}={\cal A}^{-1}$ 
\cite{HLM}. 
To probe the Fermi surface, we need to
analyse non-zero momenta $\vec k$; 
indeed the mass gap extracted from the decay
of ${\cal G}(\vec k)$ in Euclidean time initially {\sl decreases\/}
as $k\nearrow k_F$, corresponding to vacating holes in the Fermi Sea,
before rising for $k\geq k_F$ indicating particle excitations.
The resulting {\sl quasiparticle dispersion relation\/} $E(k)$ is shown in 
Fig.~\ref{fig:quasi}.
The detailed form of $E(k)$ indicates a value for $k_F\lapprox\mu$
and a Fermi velocity $v_F=\partial E/\partial k\vert_{k=k_F}\simeq0.7c$. This
is characteristic of a relativistic Fermi liquid with a repulsive interaction
between quasiparticles with parallel momenta \cite{BC}. Most importantly, there
is no evidence for a BCS gap $\Delta\not=0$; certainly $\Delta\sim\Sigma$
can be excluded. The origin of the new universality class may therefore be 
attributed to the presence of massless fermions at the Fermi surface.
The overall conclusion is that NJL$_{2+1}$ describes a relativistic thin film
gapless superfluid.

\section{A conjecture about superconductivity}

All the systems we have been able to study at high density 
(ie. those with a real measure) exhibit superfluidity. What are the prospects 
for studying the breaking of a local symmetry? We have already seen that the
only variant of TCQCD with a potentially superconducting solution, namely $N=1$
adjoint staggered flavor, is afflicted with a sign problem. The most promising 
microscopic model of high-$T_c$ superconductors, namely the Hubbard model with
on-site repulsion away from half-filling, also has an intractable sign problem
in precisely the regime of interest \cite{sugar}. 
A more familiar example in a particle physics
context is technicolor, in which condensation of
quark pairs from different representations of a gauge group 
force its dynamical breakdown; since this requires chiral fermions, in
complex represenations, a sign problem of some sort seems inevitable. An exotic
$2+1d$ example is ``$\tau_3$-QED'' \cite{DM}. This
exhibits planar superconductivity by generating an electromagnetic
photon mass by mixing with 
a second ``statistical'' photon $a_\mu$  
via a mixed Chern-Simons interaction generated
by fermion loops; for a non-zero CS coupling dynamical mass generation
through a gauge-invariant condensate $\langle \bar\psi\psi\rangle\not=0$ is
required. In
Euclidean space, however, the CS term is pure imaginary and the
resulting effective action complex; 
this also follows from the bare action since
in this case the coupling $i\bar\psi a{\!\!\!/\,}\gamma_3\gamma_5\psi$ 
implies $\{\gamma_5,D{\!\!\!\!/\,}\}\not=0$, in turn implying 
$\mbox{det}M\not=\mbox{det}M^*$.  Finally, of course, there is QCD itself.

The following
conjecture suggests itself: any system which exhibits the spontaneous
breaking of a local symmetry by a pairing
mechanism has a sign problem when formulated in terms 
of local gauge covariant degrees of freedom.

\section{Summary}

Significant progress -- both technical and psychological -- 
has been made in QCD($\mu$), resulting in the first
non-trivial LGT prediction in the $(\mu,T)$ plane. For once we have been lucky;
the high-$T$ low-$\mu$ region where simulations can probe is precisely the 
regime of direct relevance to RHIC phenomenology.  I anticipate much activity in
this area in the coming year. At $T=0$ a quantitative description of 
nuclear or quark matter regrettably seems as elusive as ever -- however
there are at least two model systems where LGT simulations can now be said 
to be doing condensed matter physics {\em ab initio\/}, ie. with matter formed
from the fundamental quanta of the theory. Two Color QCD, a
confining theory with no Fermi surface, describes the 
condensation of tightly-bound bosons, and thus resembles
superfluid $^4$He.
The NJL model, a theory without confinement but possessing a Fermi surface, 
displays unexpectedly interesting behaviour in $2+1d$ and certainly has the
potential to exhibit a fully-fledged BCS mechanism in $3+1d$, providing
a relativistic analogue of superfluid $^3$He. Simulations of NJL$_{3+1}$
will furnish non-trivial tests of model calculations of color 
superconductivity \cite{super,BR}. Ultimately, though, a microscopic
description of the superconducting state may actually {\sl require\/} a sign
problem.

\section*{Acknowledgements}

This work is partially supported by 
EU contract ERBFMRX-CT97-0122. It is my pleasure to thank 
all my collaborators,
but especially \hbox{Susan Morrison}, for their hard work and inspiration in
equal measure.

\end{document}